\documentclass[aps,prl,reprint,superscriptaddress,nobibnotes]{revtex4-1}
\usepackage{graphicx}
\usepackage{amsmath}

\begin{document}

\title{Casimir force and \textit{in-situ} surface potential measurements on nanomembranes}

\author{Daniel Garcia-Sanchez}
\author{King Yan Fong}
\author{Harish Bhaskaran}
\affiliation{Department of Electrical Engineering, Yale University, New Haven, Connecticut 06520, USA}

\author{Steve Lamoreaux}
\affiliation{Department of Physics, Yale University, New Haven, Connecticut 06520, USA}

\author{Hong X. Tang}
\affiliation{Department of Electrical Engineering, Yale University, New Haven, Connecticut 06520, USA}

\date{\today}

\begin{abstract}
We present Casimir force measurements in a sphere-plate configuration that consists of a high quality nanomembrane resonator and a millimeter sized gold coated sphere. The nanomembrane is fabricated from stoichiometric silicon nitride metallized with gold. A Kelvin probe method is used \textit{in situ} to image the surface potentials to minimize the distance-dependent residual force. Resonance-enhanced frequency-domain measurements of the nanomembrane motion allow for very high resolution measurements of the Casimir force gradient (down to a force gradient sensitivity of $3\,\mathrm{\mu N/m}$). Using this technique, the Casimir force in the range of $100\,\mathrm{nm}$ to $2{\mu m}$ is accurately measured. Experimental data thus obtained indicate that the device system in the measured range is best described with the Drude model.
\end{abstract}


\maketitle

In quantum theory the electromagnetic fields in vacuum fluctuate as a consequence of the Heisenberg uncertainty principle. It is well known that when two perfectly conducting plates of area $A$ are brought together by a distance $z$, an attractive force arises~~\cite{HBGCasimirProcAkad1948}. The interaction energy (per unit area) at zero temperature
is given by {$E(z) = \frac{\hbar c \pi^2}{720 {z^3}}$}. In the case of real metals the finite conductivity and thermal effects have to be taken into account. The corrected energy can be calculated using the Casimir-Lifshitz formalism~\cite{MBostromPRL2000}, within which the model used to describe the complex permittivity $\epsilon(\omega)$ can substantially modify the calculation of the energy, where $\omega$ is the angular frequency of the electromagnetic wave. There have been debates~\cite{MBostromPRL2000,IBrevikPRE2005,VBBezerraPRA2004} about the model for the permittivity at low frequencies: the key question was whether the TE (transverse-electrical) mode for $\omega=0$ contributes to the Casimir force. In the plasma model of free electrons, beyond the plasma frequency the metal becomes transparent, and the TE mode ($\omega=0$) contributes to the total force. In the Drude model, on the other hand, there is no contribution of the TE mode ($\omega=0$) to the total Casimir force. It has been recently reported~\cite{AOSushkovNatPhys2011} that the Drude model describes  the Casimir force in the range of $0.7$ to $7\,\mathrm{\mu m}$ with a higher accuracy, therefore excluding the plasma model in that range~\cite{AOSushkovNatPhys2011}, whereas earlier results suggest that the plasma model should apply at smaller plate separations~\cite{RSDeccaAP2005}. Thus the permittivity model at short distance remains an open question.

Broadly speaking, the Casimir force measurements can be categorized into two size regimes: macroscale and microscale. The macroscale measurement setup is usually used in the study of the Casimir force between centimeter objects~~\cite{GBressiPRL2002, SKLamoreauxPRL1997} whereas the microscale measurements utilize MEMS devices or micro-cantilevers as sensitive force transducers~\cite{UMohideenPRL1998,HBChanScience2011,HBChanPRL2001}. The macroscale measurement setup has an excellent performance at distances larger than $0.6\,\mathrm{\mu m}$ but at shorter distances, it suffers from surface contamination due to the relatively large device areas, in particular micron-scale particles. The low frequency, very compliant torsional balance involved in these measurements further limits the distance of approach, owing to environmental variations (such as seismic effects, building vibrations, etc.). In measurements involving microscale devices, particulate contamination is less of a concern. However the measurable separation of the Casimir force is also smaller due to fast scaling down of the Casimir force with reduction of interaction area. In spite of this, one can obtain a higher force sensitivity due to the miniaturized force sensor.

In this letter, we report measurements on a new Casimir Force sensor that bridges the measurement at microscale and macroscale by utilizing a nanomembrane of millimeter lateral dimensions as a sensitive force transducer. A separate millimeter sized gold coated sphere is used in a sphere-plate configuration to approach the nanomembrane. Since both surfaces have relatively large areas, a sizable Casimir force can be measured even at larger separation distances. The nanomembrane, fabricated from stoichiometric silicon nitride, retains a reasonably high quality factor even after gold metallization, therefore enables high force sensitivity. More importantly, due to the large built-in tensile stress of the stoichiometric nitride, the net stress of the bilayer remains tensile, which guarantees nanometer flatness ($3\,\mathrm{nm}$) over the whole device area ($1\,\mathrm{mm}$ $\times$ $1\,\mathrm{mm}$). The difficulty in controlling the surface flatness and particulate contamination in traditional measurement schemes is thus mitigated.

Most notable is that this new Casimir force sensor also allows for in-situ measurements of contact potentials utilizing Scanning Kelvin probe principle. It is known that, although in the case of an ideally clean conductor the surface should be equipotential, that is not usually the case in real metals~\cite{WJKimPRA2010}. The contact potential is not homogeneous along the surface and thus generates a surface potential. Such potentials may have several origins such as oxide films or adsorbed chemicals on the surface. To achieve precision Casimir force measurements, it is important to minimize the electrostatic contribution to the measured force. Usually a constant DC voltage is applied to the sphere to cancel the residual potential~\cite{UMohideenPRL1998,HBChanScience2011,HBChanPRL2001}. By scanning the metal sphere over the membrane, we are able to image the spatial distribution of the contact potential \textit{in-situ}. We show that the surface potential generates an electrostatic residual force that can not be compensated by applying a fixed DC voltage between the two plates. Instead, a separation dependent potential has to be applied in real time to cancel out the contribution from electrostatic forces. With these improvements, we have achieved unambiguous measurements of the Casimir force in the $100\,\mathrm{nm}$ -- $2\,\mathrm{\mu m}$ range.

\begin{figure}[ht]
\begin{center}
\includegraphics{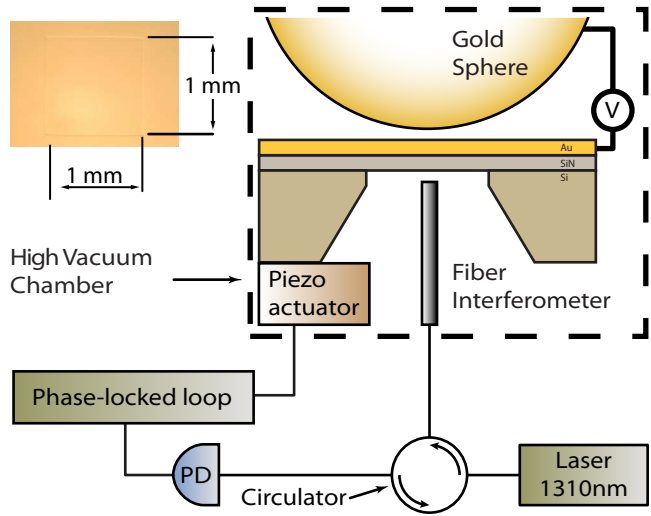}
\end{center}
\caption{Schematic of the measurement apparatus. In the inset an optical image of the gold coated membrane is shown.}\label{fig:Setup}
\end{figure}

The silicon nitride nanomembranes of $1\,\mathrm{mm}\times1\,\mathrm{mm}$ are fabricated using bulk micromachining through the handle silicon wafer. These nanomembranes were subsequently coated with $200\,\mathrm{nm}$ of Au using a \textit{e}-beam evaporator. Silicon nitride nanoresonators have been demonstrated to have very high quality factors at resonance~\cite{IWilsonRaePRL2011}. Prior to the metal coating, the mechanical quality factor of the nitride membranes exceeds 1 000 000. After metal coating, the quality factor deteriorates depending on the metal patterns on the membrane. In our Casimir studies, we choose to metallize the whole chip and fully cover the membrane so that the Casimir force is dominated by the interaction between the gold surfaces. This also allows us to apply electrical potentials to the interacting surface across the gap for electro-static characterizations. In this case, the quality factor drops to approximately 10,000 $-$ 20,000, which is still significantly higher than other types of metal resonators~\cite{DGarciaSanchezRevSciInst2011}. An image of a gold coated nanomembrane is shown in the inset in Fig.~\ref{fig:Setup}.

Our measurement setup consists of a fiber interferometer that measures the nanomembrane displacement  on one side of the membrane and a sphere on the other side. The sphere has a radius of $R = 4\,\mathrm{mm}\pm\,2.5\mathrm{\mu m}$\,\cite{RefSpherePRL} and is coated with $200\,\mathrm{nm}$ of gold. Each of the components - the membrane, the sphere, and the fiber - are individually mounted on a set of XYZ stages, driven by picomotors for coarse-alignment. An additional set of 3-axes scanning stages (PI Nanocube, 100$\mu m$ range per axis) are mounted on a sample stage to achieve sample lateral scanning and interferometer stabilization. The sphere is brought to approach the nanomebrane with a closed-loop piezo actuator with subnanometer resolution~($0.3\,\mathrm{nm}$). A schematic of the setup is illustrated in Fig.~\ref{fig:Setup}. All the components are made vacuum compatible prior to their installation in the vacuum chamber. Before the sample is introduced in the vacuum chamber, it is sealed in the cleanroom within a desiccator filled with inert gas. In order to maintain high Q, measurements were taken at pressures below $10^{-6}\,\mathrm{Torr}$, sustained by an ion-pump which eliminates mechanical vibrations. The vacuum chamber is further mounted on a damped $1500\,\mathrm{kg}$ granite table. A wood triangle is inserted between the vacuum chamber and the granite table to achieve further damping. In all our measurements, the room temperature is regulated at $20\pm 0.1^\circ\mathrm{C}$.

We conduct frequency-domain measurements of the mechanical resonator, whereby the Casimir Force gradient modifies the frequency of the nanoresonator. This technique yields better stability than static measurements. In addition, the higher Q factor also yields higher sensitivities in frequency measurement~\cite{HBChanPRL2001}. In order to track frequency shifts, the nanomembrane is driven by a piezo-actuator, and its motion is readout with the fiber interferometer. A lock-in amplifier (Zurich Instruments HF2) with a Phase-Lock-Loop (PLL) module is employed for these mesurements.

While approaching the sphere, the force experienced by the membrane has contributions from the Casimir Force as well as the electrostatic force. If we consider the first 3 terms of the Taylor expansion of
the external force $F(z)$ about the distance $z_0$, the equation of motion of the nanomembrane becomes
\begin{eqnarray}
 \ddot{z}+2\gamma\dot{z}+\omega^2_m (z-z_0)&=&\frac{F_d(z_0)}{m_{eff}}e^{i \omega_d t}+\frac{1}{m_{eff}}F'(z_0)(z-z_0) \nonumber\\
&&+\frac{1}{m_{eff}}\frac{F''(z_0)}{2} (z-z_0)^2\nonumber\\
&&+\frac{1}{m_{eff}}\frac{F'''(z_0)}{6} (z-z_0)^3\label{eq:equationmv}
\end{eqnarray}
where $\omega_m=2\pi f_m$ is the fundamental angular frequency of the membrane, $\omega_d$ the angular driving frequency, $\gamma$ is the damping coefficient, $F_d(z_0)$ is the driving force, $m_{eff}$ the effective mass of the nanomembrane and
$F'$, $F''$, and $F'''$ are the derivatives of the external force with respect to the distance $z$. Although the resonator is set to operate in the
linear regime there is some contribution from high order derivatives of the external force in the resonance frequency. The resonance frequency is modified by
\begin{eqnarray}
 \Delta f&=&-\frac{f_m}{2 k_{eff}}\left(F'(z_0)+\frac{A^2_{\mathrm{rms}}}{6}F'''(z_0)\right)\nonumber\\
 &=&-\frac{f_m}{2 k_{eff}}F'_{\mathrm{a}}(z_0)\label{eq:freqshiftforce}
\end{eqnarray}
where $F'_{\mathrm{a}}$ is the apparent force~\cite{SKLamoreauxPRA2010}, $A_{\mathrm{rms}}$ the RMS amplitude of motion of the nanomembrane and $k_{eff} = m_{eff}{\omega_0}^2$ is the spring constant. The biggest contribution comes from the first derivative of the force~\cite{MBrownHayesPRA2005} and the term with the third derivative of the force can be considered as a correction which is equivalent to the corrections for the roughness and the fluctuations of the plates that have been considered before~\cite{GLKlimchitskayaPRA1999,AOSushkovNatPhys2011}.

\begin{figure}[ht]
\begin{center}
\includegraphics{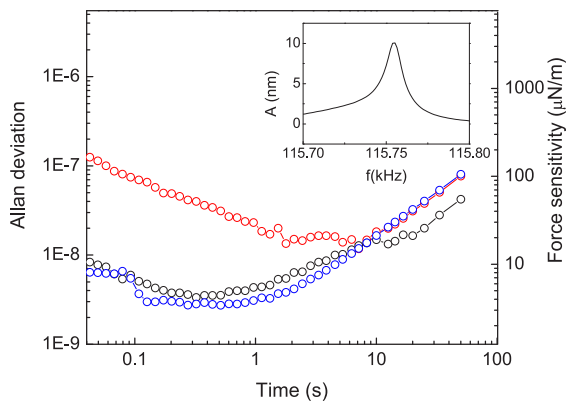}
\end{center}
\caption{Left axis shows the Allan variance vs the integration time for several drive voltages: $500\,\mathrm{\mu V}$ (red), $10\,\mathrm{mV}$
(black), and $50\,\mathrm{mV}$ (blue). The right axis shows the equivalent gradient force sensitivity. The distance between the membrane and the sphere is $1\,\mathrm{\mu m}$. Inset, mechanical resonance $Q=14\,000$.}\label{fig:AllanVariance}
\end{figure}

We first evaluate the frequency resolution by measuring the frequency fluctuations of the resonators at a fixed membrane-sphere distance ($1\,\mathrm{\mu m}$). The Allan deviation is shown in Fig.~\ref{fig:AllanVariance}, indicating that frequency resolution down to $2\cdot10^{-9}$ can be achieved with integration time of $0.4\,\textrm{s}$.
To achieve this resolution high stability of the resonance frequency is required, which is obtained by isolating the system from enviromental fluctuations. No ambient light is allowed to enter into the chamber to avoid thermal fluctuations. The temperature is stabilized within $0.1^\circ\textrm{C}$ and the granite table damps the external mechanical vibrations.
This frequency resolution allows us to measure a force gradient of $3\,\mathrm{\mu N/m}$.

Having established high frequency resolution, we approach the sphere close to the membrane resonator utilizing the closed-loop piezoactuator.
From Eq.~(\ref{eq:equationmv}) it can be found that the frequency shift can be fitted to a parabola~\cite{WJKimPRL2009}
\begin{equation}
 f^2 = f_0(z)^2-K_p(z)(V-V_m(z))^2\label{eq:equationparabola}
\end{equation}
where $V$ is the voltage applied between the membrane and the sphere. At each distance, a set of 3 voltages $V$ is applied~\cite{DGarciaSanchezRevSciInst2011}. With these measurements we can find the fitting parameters: $f_0(z)$, $K_p(z)$ and $V_m(z)$. To improve the signal to noise ratio, the measurements are repeated approaching the sphere and retracting many times.
$V_m$ represents the voltage that is required to minimize the electrostatic force given by the second term of the equation.
It is worth emphasizing that the minimizing potential $V_m$ is not necessarily constant with the distance or the relative position between the membrane and the sphere because of the nonuniform work function or contact potential across the plate. It has been suggested that such a variation of $V_m$ can cause an additional electrostatic force~\cite{WJKimPRA2010,SLamoreauxArXiv2008}.
It can be found that $K_p(z)=\epsilon_0\pi Rf^2_m/k_\mathrm{eff}(z-z_\mathrm{off})^2$
where $R$ is the radius of the sphere, $z-z_\mathrm{off}$ the absolute distance, $z_\mathrm{off}$ the offset distance, $k_\mathrm{eff}$ the effective stiffness of the membrane and $\epsilon_0$ the vacuum permittivity. The parameters $k_\mathrm{eff}$ and $z_\mathrm{off}$ can be calculated by fitting the measured $K_p(z)$ to the expected model~\cite{WJKimPRL2009}. After calibration, for each measurement, the distance is calculated from the measured $K_p$ and the calibrated parameter $k_\mathrm{eff}$. We have estimated that the error in the position is about $1\,\mathrm{nm}$ at the closest distance. The measured stiffness $k_\mathrm{eff}$ is $4000\,\mathrm{N/m}$.

In Eq.(\ref{eq:equationparabola}), the first term $f_0^2$ has contributions from the Casimir force as well as from a residual electrostatic force that cannot be canceled, given by
\begin{equation}
 \Delta f_0 = - \frac{f_m}{2 k_{\mathrm{eff}}}\left(\frac{dF_c(z)}{dz}+\frac{dF^{el}_{\mathrm{res}}(z)}{dz}\right)
\end{equation}
where $F_c$ is the Casimir force and $F^{el}_{\mathrm{res}}(z)$ is the residual electrostatic force. By considering that the surface potentials are stochastic Kim et al.~\cite{WJKimPRA2010} deduced a model for the residual electrostatic force as below:
\begin{equation}
F^{el}_{res}(z) = \pi R \epsilon_0 [(V_m(z)+V_1)^2+V^2_{\mathrm{rms}}]/z\label{eq:electrosforce}
\end{equation}
where $V_1$ and $V_{\mathrm{rms}}$ are fitting parameters. $V_{\mathrm{rms}}$ accounts for the force originated by the patches that are
smaller than the separation between the sphere and the plate.

The unambiguous measurement of the Casimir force also requires a mesurement of the surface contact potential distribution to ensure that the residual electrostatic force does not shield the Casimir force. Kelvin probe microscopes and its variations~\cite{NARobertsonClassQuanGrav2006,MNonnenmacherAPL1991,AKikukawaAPL1995} have become reliable tools to inspect the contact potential difference (CPD) distribution of a surface.
With our microscope we can also image the CPD between the sample and the sphere by scanning the sphere with respect to the membrane and simultaneously recording $V_m$. The CPD distribution is directly related to the spatial distribution of the patches on the sample. The patches on the sphere cannot be directly observed.

\begin{figure}[ht]
\begin{center}
\includegraphics{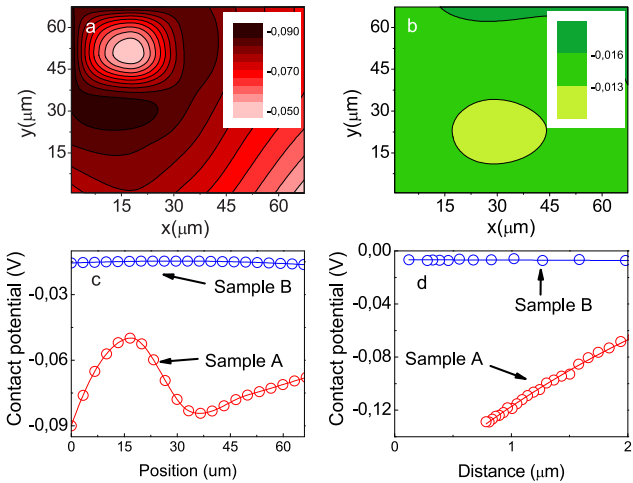}
\end{center}
\caption{(a) Measured surface potential in volts of the sample A. The distance between the membrane and the sphere
is $2\,\mathrm{\mu m}$. The data have been interpolated. (b) Measured surface potential in volts of the sample B. The distance between the membrane and the sphere
is $0.15\,\mathrm{\mu m}$. At $2\,\mathrm{\mu m}$  the surface potential variations are too small to be resolved. (c) Line scan of surface potential at $y=50\,\mathrm{\mu m}$ for the samples A and B. (d) The contact potential $V_m$ for the samples A and B vs the distance with the sphere.}\label{fig:ContactPotential}
\end{figure}

\begin{figure}[t]
\begin{center}
\includegraphics{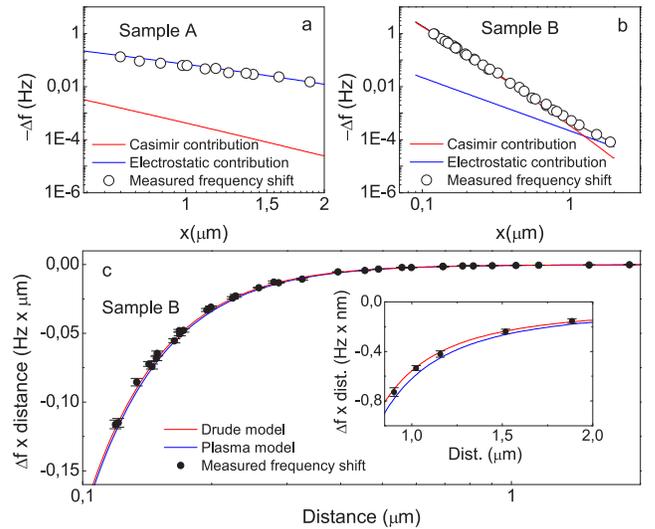}
\end{center}
\caption{(a) The measured frequency of sample A is compared with the expected frequency due to the
Casimir force and the residual electrostatic force. (b) The measured frequency of sample B is compared with the expected frequency due to the
the casimir force and the residual electrostatic force. (c) The measured frequency for the sample B is compared with the expected frequency due to the Casimir force after correcting the contribution from the electrostatic patch for both the Drude and Plasma models. The inset shows a detail of the data. The error variance of the data have several origins such as the transducer noise and the precise determination of the distance between the sphere and the  membrane.}
\label{fig:CasimirElectrostaticForce}
\end{figure}

We show the results for two membranes on the same wafer, measured by the same sphere without breaking the vacuum. The first sample labeled ``A'' has large variations of the contact potential across the surface~[see Figures~\ref{fig:ContactPotential}a and \ref{fig:ContactPotential}c]. The variations of the contact potential depend on the relative position between the sphere and the nanomembrane in $x$, $y$ and $z$.
Because of the nonuniform contact potential across the surface, the contact potential $V_m$ also largely depends on the distance between the sphere and the
nanomembrane~[see Fig.~\ref{fig:ContactPotential}d].
These variations of the contact potential create an electric field between the sphere and the nanomembrane~\cite{WJKimPRA2010}. Because the energy of this field depends on the distance between the sphere and the membrane, there is a residual electrostatic force that appears.
As mentioned above, this force can not be completely canceled by applying a voltage between the sphere and the nanomebrane~\cite{CCSpeakePRL2003,WJKimPRA2010}.
In Fig.~\ref{fig:CasimirElectrostaticForce}a we show the contributions for this sample of the Casimir force and the residual electrostatic force. Because variations of the
surface potential are large, the dominant force is electrostatic with the Casimir force a negligibly small fraction. The residual electrostatic force is well described by the model
of Eq.~(\ref{eq:electrosforce}) where $V_{\mathrm{rms}}=0.14\mathrm{V}$.

However, for the sample labeled ``B'' the spatial variation of the contact potential $V_m$ is very small~[see Figs.~\ref{fig:ContactPotential}b and \ref{fig:ContactPotential}c]. Because of the small variation of the surface contact potential, the variation of the contact potential with the distance is also small~[see Fig. \ref{fig:ContactPotential}d].
These variations are small in comparison with the variations of the sample A. As a consequence, as shown in Fig.~\ref{fig:CasimirElectrostaticForce}b the electrostatic force is negligible below $400\,\mathrm{nm}$, but it has to be taken into account above this distance and it is well described by the model
of Eq.~(\ref{eq:electrosforce}).

Hence, because of the smaller electrostatic force, Sample B is employed for studying the Casimir force. The fluctuations of the position of the membrane have to be taken into account~\cite{JLVanbreePhysica1974,SKLamoreauxRPP2005}. The origin of these fluctuations are the roughness which is about $3\,\mathrm{nm}$ and the vibrations of the membrane which have an rms amplitude of $10\,\mathrm{nm}$. The correction that has to be applied to the Casimir force and the residual electrostatic force can be calculated from Eq.~(\ref{eq:freqshiftforce}). Also the distance has to be corrected by a factor $\sqrt{1+\left(A_{\mathrm{rms}}/z\right)^2}$ because it has been extracted from the electrostatic force.

We further compare our measured results in sample B with the predictions from the Drude model~\cite{MBostromPRL2000} and the plasma model with the parameters $\omega_p=7.54\,\mathrm{eV}$ and $\gamma=0.051\,\mathrm{eV}$~\cite{AOSushkovNatPhys2011}. With the Drude model, a least squares fit shows $V_{rms}=11.6\pm 0.2\,\mathrm{mV}$ with reduced $\chi^2$ (33 degrees of freedom) of 1.07 (probability to exceed 35\%). On the other hand, a fit to  the plasma model shows $V_{rms}=8.8\pm 0.5\,\mathrm{mV}$ with a reduced $\chi^2$ of 1.7  (probability to exceed 1\%), suggesting that the Plasma model is ruled out to 99\% confidence over this distance range. Figure~\ref{fig:CasimirElectrostaticForce}c shows the measurements compared with the expected frequency shift due to the Casimir force after correcting the contribution from the electrostatic patch for both the Drude and Plasma models.

By employing high-Q nanomembranes with a force gradient sensitivity of $3\,\mathrm{\mu N}/\mathrm{m}$ we measured the Casimir force at $100\,\mathrm{nm}$-$2\,\mathrm{\mu m}$ separations. Our measurements show unambiguously that contact potentials play an important role in the precise measurement of the Casimir Force. By employing an in-situ surface potential measurements on our nanomembrane, we evaluate this uncertainty in measurements of the Casimir force. This reveals much scope for further improvements in accuracy by including methods to image such potentials on the sphere itself, which was not done in the measurements reported herein. Our data set indicates that the Drude model offers a better description of the mechanism in this range as compared to the plasma model.

This project was supported by DARPA/MTO’s Casimir  Effect Enhancement project under SPAWAR contract no. N66001-09-1-2071. H. X. Tang acknowledges a Packard Fellowship in Science and Engineering and a CAREER Grant from National Science Foundation.

\bibliography{CasimirForceMembrane}

\begin{thebibliography}{26}%
\makeatletter
\providecommand \@ifxundefined [1]{%
 \@ifx{#1\undefined}
}%
\providecommand \@ifnum [1]{%
 \ifnum #1\expandafter \@firstoftwo
 \else \expandafter \@secondoftwo
 \fi
}%
\providecommand \@ifx [1]{%
 \ifx #1\expandafter \@firstoftwo
 \else \expandafter \@secondoftwo
 \fi
}%
\providecommand \natexlab [1]{#1}%
\providecommand \enquote  [1]{``#1''}%
\providecommand \bibnamefont  [1]{#1}%
\providecommand \bibfnamefont [1]{#1}%
\providecommand \citenamefont [1]{#1}%
\providecommand \href@noop [0]{\@secondoftwo}%
\providecommand \href [0]{\begingroup \@sanitize@url \@href}%
\providecommand \@href[1]{\@@startlink{#1}\@@href}%
\providecommand \@@href[1]{\endgroup#1\@@endlink}%
\providecommand \@sanitize@url [0]{\catcode `\\12\catcode `\$12\catcode
  `\&12\catcode `\#12\catcode `\^12\catcode `\_12\catcode `\%12\relax}%
\providecommand \@@startlink[1]{}%
\providecommand \@@endlink[0]{}%
\providecommand \url  [0]{\begingroup\@sanitize@url \@url }%
\providecommand \@url [1]{\endgroup\@href {#1}{\urlprefix }}%
\providecommand \urlprefix  [0]{URL }%
\providecommand \Eprint [0]{\href }%
\providecommand \doibase [0]{http://dx.doi.org/}%
\providecommand \selectlanguage [0]{\@gobble}%
\providecommand \bibinfo  [0]{\@secondoftwo}%
\providecommand \bibfield  [0]{\@secondoftwo}%
\providecommand \translation [1]{[#1]}%
\providecommand \BibitemOpen [0]{}%
\providecommand \bibitemStop [0]{}%
\providecommand \bibitemNoStop [0]{.\EOS\space}%
\providecommand \EOS [0]{\spacefactor3000\relax}%
\providecommand \BibitemShut  [1]{\csname bibitem#1\endcsname}%
\let\auto@bib@innerbib\@empty
\bibitem [{\citenamefont {Casimir}(1948)}]{HBGCasimirProcAkad1948}%
  \BibitemOpen
  \bibfield  {author} {\bibinfo {author} {\bibfnamefont {H.~B.~G.}\
  \bibnamefont {Casimir}},\ }\href@noop {} {\bibfield  {journal} {\bibinfo
  {journal} {Proc. K. Ned. Akad. Wet.}\ }\textbf {\bibinfo {volume} {51}},\
  \bibinfo {pages} {793} (\bibinfo {year} {1948})}\BibitemShut {NoStop}%
\bibitem [{\citenamefont {Bostr\"om}\ and\ \citenamefont
  {Sernelius}(2000)}]{MBostromPRL2000}%
  \BibitemOpen
  \bibfield  {author} {\bibinfo {author} {\bibfnamefont {M.}~\bibnamefont
  {Bostr\"om}}\ and\ \bibinfo {author} {\bibfnamefont {B.~E.}\ \bibnamefont
  {Sernelius}},\ }\href {\doibase 10.1103/PhysRevLett.84.4757} {\bibfield
  {journal} {\bibinfo  {journal} {Phys. Rev. Lett.}\ }\textbf {\bibinfo
  {volume} {84}},\ \bibinfo {pages} {4757} (\bibinfo {year}
  {2000})}\BibitemShut {NoStop}%
\bibitem [{\citenamefont {Brevik}\ \emph {et~al.}(2005)\citenamefont {Brevik},
  \citenamefont {Aarseth}, \citenamefont {H\o{}ye},\ and\ \citenamefont
  {Milton}}]{IBrevikPRE2005}%
  \BibitemOpen
  \bibfield  {author} {\bibinfo {author} {\bibfnamefont {I.}~\bibnamefont
  {Brevik}}, \bibinfo {author} {\bibfnamefont {J.~B.}\ \bibnamefont {Aarseth}},
  \bibinfo {author} {\bibfnamefont {J.~S.}\ \bibnamefont {H\o{}ye}}, \ and\
  \bibinfo {author} {\bibfnamefont {K.~A.}\ \bibnamefont {Milton}},\ }\href
  {\doibase 10.1103/PhysRevE.71.056101} {\bibfield  {journal} {\bibinfo
  {journal} {Phys. Rev. E}\ }\textbf {\bibinfo {volume} {71}},\ \bibinfo
  {pages} {056101} (\bibinfo {year} {2005})}\BibitemShut {NoStop}%
\bibitem [{\citenamefont {Bezerra}\ \emph {et~al.}(2004)\citenamefont
  {Bezerra}, \citenamefont {Klimchitskaya}, \citenamefont {Mostepanenko},\ and\
  \citenamefont {Romero}}]{VBBezerraPRA2004}%
  \BibitemOpen
  \bibfield  {author} {\bibinfo {author} {\bibfnamefont {V.~B.}\ \bibnamefont
  {Bezerra}}, \bibinfo {author} {\bibfnamefont {G.~L.}\ \bibnamefont
  {Klimchitskaya}}, \bibinfo {author} {\bibfnamefont {V.~M.}\ \bibnamefont
  {Mostepanenko}}, \ and\ \bibinfo {author} {\bibfnamefont {C.}~\bibnamefont
  {Romero}},\ }\href {\doibase 10.1103/PhysRevA.69.022119} {\bibfield
  {journal} {\bibinfo  {journal} {Phys. Rev. A}\ }\textbf {\bibinfo {volume}
  {69}},\ \bibinfo {pages} {022119} (\bibinfo {year} {2004})}\BibitemShut
  {NoStop}%
\bibitem [{\citenamefont {Sushkov}\ \emph {et~al.}(2011)\citenamefont
  {Sushkov}, \citenamefont {Kim}, \citenamefont {Dalvit},\ and\ \citenamefont
  {Lamoreaux}}]{AOSushkovNatPhys2011}%
  \BibitemOpen
  \bibfield  {author} {\bibinfo {author} {\bibfnamefont {A.~O.}\ \bibnamefont
  {Sushkov}}, \bibinfo {author} {\bibfnamefont {W.~J.}\ \bibnamefont {Kim}},
  \bibinfo {author} {\bibfnamefont {D.~A.~R.}\ \bibnamefont {Dalvit}}, \ and\
  \bibinfo {author} {\bibfnamefont {S.~K.}\ \bibnamefont {Lamoreaux}},\ }\href
  {http://dx.doi.org/10.1038/nphys1909} {\bibfield  {journal} {\bibinfo
  {journal} {Nat Phys}\ }\textbf {\bibinfo {volume} {7}},\ \bibinfo {pages}
  {230} (\bibinfo {year} {2011})}\BibitemShut {NoStop}%
\bibitem [{\citenamefont {Decca}\ \emph {et~al.}(2005)\citenamefont {Decca},
  \citenamefont {López}, \citenamefont {Fischbach}, \citenamefont
  {Klimchitskaya}, \citenamefont {Krause},\ and\ \citenamefont
  {Mostepanenko}}]{RSDeccaAP2005}%
  \BibitemOpen
  \bibfield  {author} {\bibinfo {author} {\bibfnamefont {R.}~\bibnamefont
  {Decca}}, \bibinfo {author} {\bibfnamefont {D.}~\bibnamefont {López}},
  \bibinfo {author} {\bibfnamefont {E.}~\bibnamefont {Fischbach}}, \bibinfo
  {author} {\bibfnamefont {G.}~\bibnamefont {Klimchitskaya}}, \bibinfo {author}
  {\bibfnamefont {D.}~\bibnamefont {Krause}}, \ and\ \bibinfo {author}
  {\bibfnamefont {V.}~\bibnamefont {Mostepanenko}},\ }\href {\doibase DOI:
  10.1016/j.aop.2005.03.007} {\bibfield  {journal} {\bibinfo  {journal} {Annals
  of Physics}\ }\textbf {\bibinfo {volume} {318}},\ \bibinfo {pages} {37 }
  (\bibinfo {year} {2005})},\ \bibinfo {note} {special Issue}\BibitemShut
  {NoStop}%
\bibitem [{\citenamefont {Bressi}\ \emph {et~al.}(2002)\citenamefont {Bressi},
  \citenamefont {Carugno}, \citenamefont {Onofrio},\ and\ \citenamefont
  {Ruoso}}]{GBressiPRL2002}%
  \BibitemOpen
  \bibfield  {author} {\bibinfo {author} {\bibfnamefont {G.}~\bibnamefont
  {Bressi}}, \bibinfo {author} {\bibfnamefont {G.}~\bibnamefont {Carugno}},
  \bibinfo {author} {\bibfnamefont {R.}~\bibnamefont {Onofrio}}, \ and\
  \bibinfo {author} {\bibfnamefont {G.}~\bibnamefont {Ruoso}},\ }\href
  {\doibase 10.1103/PhysRevLett.88.041804} {\bibfield  {journal} {\bibinfo
  {journal} {Phys. Rev. Lett.}\ }\textbf {\bibinfo {volume} {88}},\ \bibinfo
  {pages} {041804} (\bibinfo {year} {2002})}\BibitemShut {NoStop}%
\bibitem [{\citenamefont {Lamoreaux}(1997)}]{SKLamoreauxPRL1997}%
  \BibitemOpen
  \bibfield  {author} {\bibinfo {author} {\bibfnamefont {S.~K.}\ \bibnamefont
  {Lamoreaux}},\ }\href {\doibase 10.1103/PhysRevLett.78.5} {\bibfield
  {journal} {\bibinfo  {journal} {Phys. Rev. Lett.}\ }\textbf {\bibinfo
  {volume} {78}},\ \bibinfo {pages} {5} (\bibinfo {year} {1997})}\BibitemShut
  {NoStop}%
\bibitem [{\citenamefont {Mohideen}\ and\ \citenamefont
  {Roy}(1998)}]{UMohideenPRL1998}%
  \BibitemOpen
  \bibfield  {author} {\bibinfo {author} {\bibfnamefont {U.}~\bibnamefont
  {Mohideen}}\ and\ \bibinfo {author} {\bibfnamefont {A.}~\bibnamefont {Roy}},\
  }\href {\doibase 10.1103/PhysRevLett.81.4549} {\bibfield  {journal} {\bibinfo
   {journal} {Phys. Rev. Lett.}\ }\textbf {\bibinfo {volume} {81}},\ \bibinfo
  {pages} {4549} (\bibinfo {year} {1998})}\BibitemShut {NoStop}%
\bibitem [{\citenamefont {Chan}\ \emph
  {et~al.}(2001{\natexlab{a}})\citenamefont {Chan}, \citenamefont {Aksyuk},
  \citenamefont {Kleiman}, \citenamefont {Bishop},\ and\ \citenamefont
  {Capasso}}]{HBChanScience2011}%
  \BibitemOpen
  \bibfield  {author} {\bibinfo {author} {\bibfnamefont {H.~B.}\ \bibnamefont
  {Chan}}, \bibinfo {author} {\bibfnamefont {V.~A.}\ \bibnamefont {Aksyuk}},
  \bibinfo {author} {\bibfnamefont {R.~N.}\ \bibnamefont {Kleiman}}, \bibinfo
  {author} {\bibfnamefont {D.~J.}\ \bibnamefont {Bishop}}, \ and\ \bibinfo
  {author} {\bibfnamefont {F.}~\bibnamefont {Capasso}},\ }\href {\doibase
  10.1126/science.1057984} {\bibfield  {journal} {\bibinfo  {journal}
  {Science}\ }\textbf {\bibinfo {volume} {291}},\ \bibinfo {pages} {1941}
  (\bibinfo {year} {2001}{\natexlab{a}})}\BibitemShut {NoStop}%
\bibitem [{\citenamefont {Chan}\ \emph
  {et~al.}(2001{\natexlab{b}})\citenamefont {Chan}, \citenamefont {Aksyuk},
  \citenamefont {Kleiman}, \citenamefont {Bishop},\ and\ \citenamefont
  {Capasso}}]{HBChanPRL2001}%
  \BibitemOpen
  \bibfield  {author} {\bibinfo {author} {\bibfnamefont {H.~B.}\ \bibnamefont
  {Chan}}, \bibinfo {author} {\bibfnamefont {V.~A.}\ \bibnamefont {Aksyuk}},
  \bibinfo {author} {\bibfnamefont {R.~N.}\ \bibnamefont {Kleiman}}, \bibinfo
  {author} {\bibfnamefont {D.~J.}\ \bibnamefont {Bishop}}, \ and\ \bibinfo
  {author} {\bibfnamefont {F.}~\bibnamefont {Capasso}},\ }\href {\doibase
  10.1103/PhysRevLett.87.211801} {\bibfield  {journal} {\bibinfo  {journal}
  {Phys. Rev. Lett.}\ }\textbf {\bibinfo {volume} {87}},\ \bibinfo {pages}
  {211801} (\bibinfo {year} {2001}{\natexlab{b}})}\BibitemShut {NoStop}%
\bibitem [{\citenamefont {Kim}\ \emph {et~al.}(2010)\citenamefont {Kim},
  \citenamefont {Sushkov}, \citenamefont {Dalvit},\ and\ \citenamefont
  {Lamoreaux}}]{WJKimPRA2010}%
  \BibitemOpen
  \bibfield  {author} {\bibinfo {author} {\bibfnamefont {W.~J.}\ \bibnamefont
  {Kim}}, \bibinfo {author} {\bibfnamefont {A.~O.}\ \bibnamefont {Sushkov}},
  \bibinfo {author} {\bibfnamefont {D.~A.~R.}\ \bibnamefont {Dalvit}}, \ and\
  \bibinfo {author} {\bibfnamefont {S.~K.}\ \bibnamefont {Lamoreaux}},\ }\href
  {\doibase 10.1103/PhysRevA.81.022505} {\bibfield  {journal} {\bibinfo
  {journal} {Phys. Rev. A}\ }\textbf {\bibinfo {volume} {81}},\ \bibinfo
  {pages} {022505} (\bibinfo {year} {2010})}\BibitemShut {NoStop}%
\bibitem [{\citenamefont {Wilson-Rae}\ \emph {et~al.}(2011)\citenamefont
  {Wilson-Rae}, \citenamefont {Barton}, \citenamefont {Verbridge},
  \citenamefont {Southworth}, \citenamefont {Ilic}, \citenamefont {Craighead},\
  and\ \citenamefont {Parpia}}]{IWilsonRaePRL2011}%
  \BibitemOpen
  \bibfield  {author} {\bibinfo {author} {\bibfnamefont {I.}~\bibnamefont
  {Wilson-Rae}}, \bibinfo {author} {\bibfnamefont {R.~A.}\ \bibnamefont
  {Barton}}, \bibinfo {author} {\bibfnamefont {S.~S.}\ \bibnamefont
  {Verbridge}}, \bibinfo {author} {\bibfnamefont {D.~R.}\ \bibnamefont
  {Southworth}}, \bibinfo {author} {\bibfnamefont {B.}~\bibnamefont {Ilic}},
  \bibinfo {author} {\bibfnamefont {H.~G.}\ \bibnamefont {Craighead}}, \ and\
  \bibinfo {author} {\bibfnamefont {J.~M.}\ \bibnamefont {Parpia}},\ }\href
  {\doibase 10.1103/PhysRevLett.106.047205} {\bibfield  {journal} {\bibinfo
  {journal} {Phys. Rev. Lett.}\ }\textbf {\bibinfo {volume} {106}},\ \bibinfo
  {pages} {047205} (\bibinfo {year} {2011})}\BibitemShut {NoStop}%
\bibitem [{\citenamefont {{Garcia-sanchez}}\ \emph {et~al.}()\citenamefont
  {{Garcia-sanchez}}, \citenamefont {Fong}, \citenamefont {Bhaskaran},
  \citenamefont {Lamoreaux},\ and\ \citenamefont
  {Tang}}]{DGarciaSanchezRevSciInst2011}%
  \BibitemOpen
  \bibfield  {author} {\bibinfo {author} {\bibfnamefont {D.}~\bibnamefont
  {{Garcia-sanchez}}}, \bibinfo {author} {\bibfnamefont {K.~Y.}\ \bibnamefont
  {Fong}}, \bibinfo {author} {\bibfnamefont {H.}~\bibnamefont {Bhaskaran}},
  \bibinfo {author} {\bibfnamefont {S.}~\bibnamefont {Lamoreaux}}, \ and\
  \bibinfo {author} {\bibfnamefont {H.~X.}\ \bibnamefont {Tang}},\ }\href@noop
  {} {}\bibinfo {note} {(To be published)}\BibitemShut {NoStop}%
\bibitem [{Ref()}]{RefSpherePRL}%
  \BibitemOpen
  \href@noop {} {}\bibinfo {note} {The sphere is a fused silica ball lens from
  Edmund Optics with ref. NT67-388.}\BibitemShut {Stop}%
\bibitem [{\citenamefont {Lamoreaux}(2010)}]{SKLamoreauxPRA2010}%
  \BibitemOpen
  \bibfield  {author} {\bibinfo {author} {\bibfnamefont {S.~K.}\ \bibnamefont
  {Lamoreaux}},\ }\href {\doibase 10.1103/PhysRevA.82.024102} {\bibfield
  {journal} {\bibinfo  {journal} {Phys. Rev. A}\ }\textbf {\bibinfo {volume}
  {82}},\ \bibinfo {pages} {024102} (\bibinfo {year} {2010})}\BibitemShut
  {NoStop}%
\bibitem [{\citenamefont {Brown-Hayes}\ \emph {et~al.}(2005)\citenamefont
  {Brown-Hayes}, \citenamefont {Dalvit}, \citenamefont {Mazzitelli},
  \citenamefont {Kim},\ and\ \citenamefont {Onofrio}}]{MBrownHayesPRA2005}%
  \BibitemOpen
  \bibfield  {author} {\bibinfo {author} {\bibfnamefont {M.}~\bibnamefont
  {Brown-Hayes}}, \bibinfo {author} {\bibfnamefont {D.~A.~R.}\ \bibnamefont
  {Dalvit}}, \bibinfo {author} {\bibfnamefont {F.~D.}\ \bibnamefont
  {Mazzitelli}}, \bibinfo {author} {\bibfnamefont {W.~J.}\ \bibnamefont {Kim}},
  \ and\ \bibinfo {author} {\bibfnamefont {R.}~\bibnamefont {Onofrio}},\ }\href
  {\doibase 10.1103/PhysRevA.72.052102} {\bibfield  {journal} {\bibinfo
  {journal} {Phys. Rev. A}\ }\textbf {\bibinfo {volume} {72}},\ \bibinfo
  {pages} {052102} (\bibinfo {year} {2005})}\BibitemShut {NoStop}%
\bibitem [{\citenamefont {Klimchitskaya}\ \emph {et~al.}(1999)\citenamefont
  {Klimchitskaya}, \citenamefont {Roy}, \citenamefont {Mohideen},\ and\
  \citenamefont {Mostepanenko}}]{GLKlimchitskayaPRA1999}%
  \BibitemOpen
  \bibfield  {author} {\bibinfo {author} {\bibfnamefont {G.~L.}\ \bibnamefont
  {Klimchitskaya}}, \bibinfo {author} {\bibfnamefont {A.}~\bibnamefont {Roy}},
  \bibinfo {author} {\bibfnamefont {U.}~\bibnamefont {Mohideen}}, \ and\
  \bibinfo {author} {\bibfnamefont {V.~M.}\ \bibnamefont {Mostepanenko}},\
  }\href {\doibase 10.1103/PhysRevA.60.3487} {\bibfield  {journal} {\bibinfo
  {journal} {Phys. Rev. A}\ }\textbf {\bibinfo {volume} {60}},\ \bibinfo
  {pages} {3487} (\bibinfo {year} {1999})}\BibitemShut {NoStop}%
\bibitem [{\citenamefont {Kim}\ \emph {et~al.}(2009)\citenamefont {Kim},
  \citenamefont {Sushkov}, \citenamefont {Dalvit},\ and\ \citenamefont
  {Lamoreaux}}]{WJKimPRL2009}%
  \BibitemOpen
  \bibfield  {author} {\bibinfo {author} {\bibfnamefont {W.~J.}\ \bibnamefont
  {Kim}}, \bibinfo {author} {\bibfnamefont {A.~O.}\ \bibnamefont {Sushkov}},
  \bibinfo {author} {\bibfnamefont {D.~A.~R.}\ \bibnamefont {Dalvit}}, \ and\
  \bibinfo {author} {\bibfnamefont {S.~K.}\ \bibnamefont {Lamoreaux}},\ }\href
  {\doibase 10.1103/PhysRevLett.103.060401} {\bibfield  {journal} {\bibinfo
  {journal} {Phys. Rev. Lett.}\ }\textbf {\bibinfo {volume} {103}},\ \bibinfo
  {pages} {060401} (\bibinfo {year} {2009})}\BibitemShut {NoStop}%
\bibitem [{\citenamefont {Lamoreaux}(2008)}]{SLamoreauxArXiv2008}%
  \BibitemOpen
  \bibfield  {author} {\bibinfo {author} {\bibfnamefont {S.}~\bibnamefont
  {Lamoreaux}},\ }\href@noop {} {\bibfield  {journal} {\bibinfo  {journal}
  {arXiv:0808.0885}\ } (\bibinfo {year} {2008})}\BibitemShut {NoStop}%
\bibitem [{\citenamefont {Robertson}\ \emph {et~al.}(2006)\citenamefont
  {Robertson}, \citenamefont {Blackwood}, \citenamefont {Buchman},
  \citenamefont {Byer}, \citenamefont {Camp}, \citenamefont {Gill},
  \citenamefont {Hanson}, \citenamefont {Williams},\ and\ \citenamefont
  {Zhou}}]{NARobertsonClassQuanGrav2006}%
  \BibitemOpen
  \bibfield  {author} {\bibinfo {author} {\bibfnamefont {N.~A.}\ \bibnamefont
  {Robertson}}, \bibinfo {author} {\bibfnamefont {J.~R.}\ \bibnamefont
  {Blackwood}}, \bibinfo {author} {\bibfnamefont {S.}~\bibnamefont {Buchman}},
  \bibinfo {author} {\bibfnamefont {R.~L.}\ \bibnamefont {Byer}}, \bibinfo
  {author} {\bibfnamefont {J.}~\bibnamefont {Camp}}, \bibinfo {author}
  {\bibfnamefont {D.}~\bibnamefont {Gill}}, \bibinfo {author} {\bibfnamefont
  {J.}~\bibnamefont {Hanson}}, \bibinfo {author} {\bibfnamefont
  {S.}~\bibnamefont {Williams}}, \ and\ \bibinfo {author} {\bibfnamefont
  {P.}~\bibnamefont {Zhou}},\ }\href
  {http://stacks.iop.org/0264-9381/23/i=7/a=026} {\bibfield  {journal}
  {\bibinfo  {journal} {Classical and Quantum Gravity}\ }\textbf {\bibinfo
  {volume} {23}},\ \bibinfo {pages} {2665} (\bibinfo {year}
  {2006})}\BibitemShut {NoStop}%
\bibitem [{\citenamefont {Nonnenmacher}\ \emph {et~al.}(1991)\citenamefont
  {Nonnenmacher}, \citenamefont {O'Boyle},\ and\ \citenamefont
  {Wickramasinghe}}]{MNonnenmacherAPL1991}%
  \BibitemOpen
  \bibfield  {author} {\bibinfo {author} {\bibfnamefont {M.}~\bibnamefont
  {Nonnenmacher}}, \bibinfo {author} {\bibfnamefont {M.~P.}\ \bibnamefont
  {O'Boyle}}, \ and\ \bibinfo {author} {\bibfnamefont {H.~K.}\ \bibnamefont
  {Wickramasinghe}},\ }\href {\doibase 10.1063/1.105227} {\bibfield  {journal}
  {\bibinfo  {journal} {Applied Physics Letters}\ }\textbf {\bibinfo {volume}
  {58}},\ \bibinfo {pages} {2921} (\bibinfo {year} {1991})}\BibitemShut
  {NoStop}%
\bibitem [{\citenamefont {Kikukawa}\ \emph {et~al.}(1995)\citenamefont
  {Kikukawa}, \citenamefont {Hosaka},\ and\ \citenamefont
  {Imura}}]{AKikukawaAPL1995}%
  \BibitemOpen
  \bibfield  {author} {\bibinfo {author} {\bibfnamefont {A.}~\bibnamefont
  {Kikukawa}}, \bibinfo {author} {\bibfnamefont {S.}~\bibnamefont {Hosaka}}, \
  and\ \bibinfo {author} {\bibfnamefont {R.}~\bibnamefont {Imura}},\ }\href
  {\doibase 10.1063/1.113780} {\bibfield  {journal} {\bibinfo  {journal}
  {Applied Physics Letters}\ }\textbf {\bibinfo {volume} {66}},\ \bibinfo
  {pages} {3510} (\bibinfo {year} {1995})}\BibitemShut {NoStop}%
\bibitem [{\citenamefont {Speake}\ and\ \citenamefont
  {Trenkel}(2003)}]{CCSpeakePRL2003}%
  \BibitemOpen
  \bibfield  {author} {\bibinfo {author} {\bibfnamefont {C.~C.}\ \bibnamefont
  {Speake}}\ and\ \bibinfo {author} {\bibfnamefont {C.}~\bibnamefont
  {Trenkel}},\ }\href {\doibase 10.1103/PhysRevLett.90.160403} {\bibfield
  {journal} {\bibinfo  {journal} {Phys. Rev. Lett.}\ }\textbf {\bibinfo
  {volume} {90}},\ \bibinfo {pages} {160403} (\bibinfo {year}
  {2003})}\BibitemShut {NoStop}%
\bibitem [{\citenamefont {Vanbree}\ \emph {et~al.}(1974)\citenamefont
  {Vanbree}, \citenamefont {Poulis}, \citenamefont {Verhaar},\ and\
  \citenamefont {Schram}}]{JLVanbreePhysica1974}%
  \BibitemOpen
  \bibfield  {author} {\bibinfo {author} {\bibfnamefont {J.~L.~M.}\
  \bibnamefont {Vanbree}}, \bibinfo {author} {\bibfnamefont {J.~A.}\
  \bibnamefont {Poulis}}, \bibinfo {author} {\bibfnamefont {B.~J.}\
  \bibnamefont {Verhaar}}, \ and\ \bibinfo {author} {\bibfnamefont
  {K.}~\bibnamefont {Schram}},\ }\href@noop {} {\bibfield  {journal} {\bibinfo
  {journal} {Physica}\ }\textbf {\bibinfo {volume} {78}},\ \bibinfo {pages}
  {187} (\bibinfo {year} {1974})}\BibitemShut {NoStop}%
\bibitem [{\citenamefont {Lamoreaux}(2005)}]{SKLamoreauxRPP2005}%
  \BibitemOpen
  \bibfield  {author} {\bibinfo {author} {\bibfnamefont {S.~K.}\ \bibnamefont
  {Lamoreaux}},\ }\href {http://stacks.iop.org/0034-4885/68/i=1/a=R04}
  {\bibfield  {journal} {\bibinfo  {journal} {Reports on Progress in Physics}\
  }\textbf {\bibinfo {volume} {68}},\ \bibinfo {pages} {201} (\bibinfo {year}
  {2005})}\BibitemShut {NoStop}%
\end{thebibliography}%

\end{document}